\documentclass[conference]{IEEEtran}
\IEEEoverridecommandlockouts
\usepackage{cite}
\usepackage{amsmath,amssymb,amsfonts}
\usepackage{algorithmic}
\usepackage{graphicx}
\usepackage{textcomp}
\usepackage{xcolor}

\usepackage{tikz}
\usetikzlibrary{shapes,
	automata,
	arrows,
	chains,
	matrix,
	backgrounds,
	fit,
	patterns,
	decorations.markings, 
	svg.path, 
	shapes.multipart, 
	shapes.symbols,
	shapes.geometric,
	external, 
	shadows, 
	positioning, 
	calc, 
	decorations, 
	scopes,
	bending}
\usetikzlibrary{external}
\usepackage{pgfplots}
\pgfplotsset{compat=1.7}
\usepackage{pgfplotstable}
\usepackage{enumitem}
\usepackage{siunitx}
\usetikzlibrary{arrows, shapes, positioning, shapes.geometric, decorations.markings, calc, fit, fadings, shadows}
\usetikzlibrary{scopes}
\usetikzlibrary{bending}
\usetikzlibrary{arrows.meta}
\usepackage[export]{adjustbox}
\usepackage{makecell}
\usepackage{booktabs}
\usepackage{xcolor}

\usepackage{comment}
\def\BibTeX{{\rm B\kern-.05em{\sc i\kern-.025em b}\kern-.08em
    T\kern-.1667em\lower.7ex\hbox{E}\kern-.125emX}}
\begin{document}

\title{An Application of Scenario Exploration to Find New Scenarios for the Development and Testing of Automated Driving Systems in Urban Scenarios}

\author{
  Barbara Schütt\IEEEauthorrefmark{1},
  Marc Heinrich\IEEEauthorrefmark{1},
  Sonja Marahrens\IEEEauthorrefmark{2},
  J. Marius Zöllner\IEEEauthorrefmark{1}, 
  Eric Sax\IEEEauthorrefmark{1}
  \\
  \IEEEauthorblockA{%
    \IEEEauthorrefmark{1}FZI Research Center for Information Technology, Karlsruhe, Germany\\
    \{schuett, heinrich, zoellner, sax\}@fzi.de
  }
  \IEEEauthorblockA{%
    \IEEEauthorrefmark{2}IPG Automotive GmbH, Karlsruhe, Germany\\
    sonja.marahrens@ipg-automotive.com
  }
  }
\maketitle

\begin{abstract}
Verification and validation are major challenges for developing automated driving systems.
A concept that gets more and more recognized for testing in automated driving is scenario-based testing. 
However, it introduces the problem of what scenarios are relevant for testing and which are not. 
This work aims to find relevant, interesting, or critical parameter sets within logical scenarios by utilizing Bayes optimization and Gaussian processes.
The parameter optimization is done by comparing and evaluating six different metrics in two urban intersection scenarios. 
Finally, a list of ideas this work leads to and should be investigated further is presented.
\end{abstract}


\section{\uppercase{Introduction}}
The development of semi-automated, automated and autonomous vehicles has  played an important role in the software and hardware departments of automotive manufacturers during recent years.
The consulting company Gartner has already anticipated \textit{autonomous things} as a ''hot topic'' several times in previous years and is now going one step further. 
The current report ''Top 10 Strategic Technology Trends for 2022'' predicts \textit{autonomic systems} as a main area of interest: systems that can not only make autonomous decisions, but additionally are able to adapt and change their behavior according to the environment \cite{Gartner}.

One type of autonomous or autonomic system are automated vehicles.
A major challenge besides their development is to ensure that the system is sufficiently safe and can be approved and permitted on public roads.
A widely discussed testing approach is scenario-based testing:
According to \cite{otten2018automated}, one goal is to take realistic field trial test drives into simulation environments, where predefined scenarios often serve as a basis for the derivation of relevant test cases in automated assessment and, thus, reducing the needed amount of real test drives.
Moreover, the proper representation and usage of scenarios during the development process support a seamless development and testing of automated driving functions, as well as the specification of requirements and automated derivation of test cases \cite{bach2016model}.
However, the specification of scenarios can include parameter ranges, where only a sub-set of these ranges might bring insight into the performance of an automated driving function or hold critical scenarios.
Additionally, introducing new parameters or parameter ranges in a scenario increases the number of scenarios exponentially.

\subsection*{Novelty and Main Contribution to the State of the Art}
The novelty and main contribution of this paper is a parameter evaluation for finding challenging and critical scenario parameters in predefined parameter ranges. 
Thus, we
\begin{itemize}
    \item optimize the parameters for different intersection scenarios with different criticality metrics to find interesting scenarios and to save simulation time, and
    \item use the information gained by the optimization process to further assess these parameters and their meaning for redefining parameter ranges and the evaluation and assessment of an automated driving function.
\end{itemize}

\subsection*{Structure}
The paper is structured as follows: 
Section~\ref{sec:related_work} briefly introduces topics related to this work, e.g., scenario abstraction levels or Bayes optimization. 
Section~\ref{sec:scenario_exploration} explains the proposed exploration algorithm and setup, including the simulation environment setup and evaluation.
In section~\ref{sec:conclusion}, we conclude and give a short overview of possible future work.

\section{\uppercase{Related Work}}
\label{sec:related_work}
In the context of this work, the terms \textit{scenario} and \textit{scene} are used as summarized by \cite{steimle2021consistent}. 
A scene is a snapshots of a traffic constellation.
A scenario is a sequence of scenes and describes the temporal development of the behavior of different actors within this sequence.

 Finding new scenarios is a relevant step for defining new test cases to assess an automated driving system's safety.
\cite{bussler2020application} use evolutionary learning to find relevant parameter sets within logical scenarios and utilize Euclidean distance and time-to-collision for their fitness evaluation to find more critical scenarios.
Another approach proposed by \cite{baumann2021automatic} uses reinforcement learning combined with the metrics headway and time-to-collision to gain new test cases.
Additionally, \cite{abeysirigoonawardena2019generating} use Bayes Optimization and Euclidean distance to generate training scenarios for a driving function to learn to avoid pedestrians by reinforcement learning. 
However, their work does not produce a scenario set suitable for testing since their approach always uses the current state of the driving function which changes over the course of the experiments.
Further, there are other approaches to find new scenarios, e.g., extracting scenarios from recorded data sets as shown by \cite{king2021capturing} and \cite{zofka2015data} or by experts planning and designing scenarios from scratch.

\subsection{Scenario Abstraction Levels}
\cite{menzel2018scenarios} suggest three abstraction levels for scenario representation.
The most abstract level of scenario representation is called \textit{functional} and describes a scenario via linguistic notation using natural, non-structured language terminology.
The main goal for this level is to create scenarios that are easily understandable and open for discussion between experts.
It describes the base road network and all actors with their maneuvers, such as a right-turning vehicle or road crossing cyclist.
The next abstraction level is the \textit{logical} level and refines the representation of functional scenarios by a detailed representation with the help of state-space variables.
These variables or parameters can, for instance, be ranges for road width, vehicle positions, and their speed, or time and weather conditions.
%The parameters are described with parameter ranges, which may include a probability distribution.
The most detailed level is called \textit{concrete} and describes operating scenarios with concrete values for each parameter in the parameter space.
Therefore, one logical scenario can yield many concrete scenarios, depending on the number of variables, size of parameter ranges, and step size for these ranges.

\subsection{Bayes Optimization and Gaussian Process}
Bayes optimization (BO) proceeds by maintaining a global statistical model of a given objective function $f(\mathbf{x})$ iteratively and consists of two main steps \cite{bayes_opt_review}:
The first step is the \textbf{Gaussian process}, which is used to represent the predicted mean $\mu_t(\mathbf{x})$ and the uncertainty $\sigma_t(\mathbf{x})$ for each point $\mathbf{x}$ of the input space, with the given set of observations $\mathcal{D}_{1:t}=\{(\mathbf{x}_1, y_1), (\mathbf{x}_2, y_2), ...(\mathbf{x}_t, y_t)\}$, where $\mathbf{x}_t$ is the process input and $y_t$ the corresponding output at time $t$.
After that, an \textbf{Acquisition function} is used to evaluate the beliefs about the objective function regarding the input space, based on the predicted mean $\mu_t(\mathbf{x})$ and uncertainty and chooses the most promising setting $\sigma_t(\mathbf{x})$.
 
\subsection{Scenario Metrics}
Scenario metrics are used to assess the quality of a scenario regarding the aspect that needs to be evaluated.
According to \cite{schutt2021taxonomy}, scenario quality can be assessed at three different levels of resolution: nanoscopic (a scenario segment is evaluated, e.g., a single time step), microscopic (a complete scenario is evaluated, e.g., one concrete scenario), and macroscopic (a set of scenarios is evaluated, e.g., a logical scenario).
Before a metric for the evaluation process is chosen, the usage, goals, and purpose of a scenario need to be clear, e.g., \cite{schonemann2018scenario} propose a hazard analysis and risk evaluation to determine safety goals and show their approach on the example of a valet parking system.
The formulated safety goals can be used in following steps to choose the metrics for scenario evaluation or to determine the performance of an automated driving system concerning its requirements.

\subsection{Simulation Tools}
Commercial tools for automotive simulation among others are available from dSPACE \cite{dSPACE},  and IPG \cite{carmaker}.
Both simulation tools provide modules for map and scenario creation, sensor models and dynamic models, to name some examples. 
A further tool is Carla, an open-source simulator with a growing community and based on the game engine Unreal \cite{Dosovitskiy17}.
It offers several additional modules, e.g., a scenario tool which includes its own scenario format, a graphical tool for creating scenarios, a ROS-bridge, and SUMO support.
SUMO is an open-source software tool for modeling microscopic traffic simulation from DLR \cite{SUMO2018}.
It specializes on big scale of traffic simulation and can be used for evaluating traffic lights cycles, evaluation of emissions (noise, pollutants), traffic forecast, and many others.

\section{\uppercase{Directed Scenario Exploration}}
\label{sec:scenario_exploration}

\subsection{Optimization Setup}
\label{sec:opt_steup}
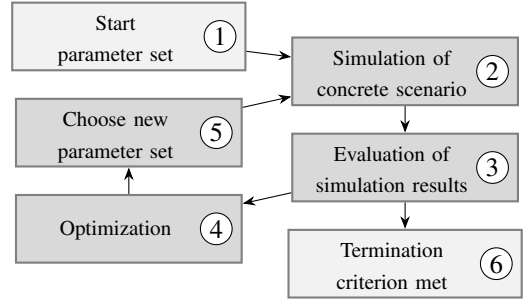
\begin{figure}[tbh]
    \centering
    \begin{adjustbox}{center,max width=\textwidth,margin*=0ex 0ex 0ex 1ex}
		\tikzsetnextfilename{simulation_flow}
\begin{tikzpicture}[
textbox/.style={rectangle,minimum height=0.9cm,minimum width=3cm,draw=black!50,fill=black!15,thick, inner sep=5pt},
textbox2/.style={rectangle,minimum height=0.9cm,minimum width=3.1cm,draw=black!50,fill=black!5,thick, inner sep=5pt},
invisibox/.style={rectangle, inner sep=5pt}
]

\node[textbox2, anchor=south, align=center] (begin) {};
\node[invisibox, anchor=north, align=center] at  ($(begin.north) + (down:0pt) + (left:5pt)$) (begin_t) {\footnotesize Start  \\ \footnotesize parameter set};
\node[draw,circle,minimum size=12pt,inner sep=0pt,draw=black,fill=white] at ($(begin.east) + (up:0pt) + (left:10pt)$) (1) {1};
\node[textbox, anchor=north, align=center] at  ($(begin.east) + (up:0pt) + (right:60pt)$) (simulation){};
\node[invisibox, anchor=north, align=center] at  ($(simulation.north) + (down:0pt) + (left:5pt)$) (simulation_t) {\footnotesize Simulation of \\ \footnotesize concrete scenario};
\node[draw,circle,minimum size=12pt,inner sep=0pt,draw=black,fill=white] at ($(simulation.east) + (up:0pt) + (left:10pt)$) (2) {2};

\node[textbox, anchor=north, align=center] at  ($(simulation.south) + (down:10pt) + (right:0pt)$) (evaluation){};
\node[invisibox, anchor=north, align=center] at  ($(evaluation.north) + (down:0pt) + (left:5pt)$) (evaluation_t) {\footnotesize Evaluation of \\ \footnotesize simulation results};
\node[draw,circle,minimum size=12pt,inner sep=0pt,draw=black,fill=white] at ($(evaluation.east) + (up:0pt) + (left:10pt)$) (3) {3};

\node[textbox, anchor=north, align=center] at  ($(begin.south) + (down:10pt) + (left:0pt)$) (parameter){};
\node[invisibox, anchor=north, align=center] at  ($(parameter.north) + (down:0pt) + (left:5pt)$) (parameter_t) {\footnotesize Choose new \\ \footnotesize parameter set};
\node[draw,circle,minimum size=12pt,inner sep=0pt,draw=black,fill=white] at ($(parameter.east) + (up:0pt) + (left:10pt)$) (5) {5};

\node[textbox, anchor=north, align=center] at  ($(parameter.south) + (down:10pt) + (left:0pt)$) (optimization){};
\node[invisibox, anchor=north, align=center] at  ($(optimization.north) + (down:5pt) + (left:5pt)$) (optimization_t) {\footnotesize Optimization};
\node[draw,circle,minimum size=12pt,inner sep=0pt,draw=black,fill=white] at ($(optimization.east) + (up:0pt) + (left:10pt)$) (4) {4};

\node[textbox2, anchor=north, align=center] at  ($(evaluation.south) + (down:10pt) + (right:0pt)$) (end){};
\node[invisibox, anchor=north, align=center] at  ($(end.north) + (down:0pt) + (left:5pt)$) (end_t) {\footnotesize Termination \\ \footnotesize criterion met};
\node[draw,circle,minimum size=12pt,inner sep=0pt,draw=black,fill=white] at ($(end.east) + (up:0pt) + (left:10pt)$) (6) {6};

\draw[arrows = {-Stealth[scale=1]}] (begin) -- (simulation);

%\draw[arrows = {-Stealth[scale=1]}] (simulation) ..controls  +(2.5,0.0).. (evaluation);
\draw[arrows = {-Stealth[scale=1]}] (simulation) -- (evaluation);
%\draw[arrows = {-Stealth[scale=1]}] (evaluation) ..controls  +(0.0,-1.0).. (optimization);
\draw[arrows = {-Stealth[scale=1]}] (evaluation) -- (optimization);

%\draw[arrows = {-Stealth[scale=1]}] (optimization) ..controls  +(-2.4,0.0).. (parameter);
\draw[arrows = {-Stealth[scale=1]}] (optimization) -- (parameter);
%\draw[arrows = {-Stealth[scale=1]}] (parameter) ..controls  +(0.0,1.2).. (simulation);
\draw[arrows = {-Stealth[scale=1]}] (parameter) -- (simulation);
\draw[arrows = {-Stealth[scale=1]}] (evaluation) -- (end);
\end{tikzpicture}
	\end{adjustbox}
    \caption{Optimization workflow}
    \label{fig:optimization}
\end{figure}
The optimization is an iterative process and is outlined in Fig.~\ref{fig:optimization}.
First, a start parameter set is selected (1), and simulated as summarized in step (2).
The results are evaluated (3), a new parameter set is chosen (5) by the optimization algorithm (4), and it is simulated again (2).
This step is repeated until a termination criterion is met (6).
The open-source project common Bayesian optimization library (COMBO) is employed in the experiments since it offers Bayesian optimization that uses automatic hyperparameter tuning, Thompson sampling as a method of picking the next best candidate, and random feature maps for better performance \cite{ueno2016combo}.
Throughout this work, Bayesian optimization and Gaussian processes were used. 
However, other optimization algorithms might be used since the focus of this work does not lie on the optimization itself.

\subsection{Simulator Setup}
\begin{figure*}[tbh]
    \centering
    \begin{adjustbox}{center,width=\linewidth,margin*=0ex 0ex 0ex 1ex}
		\input{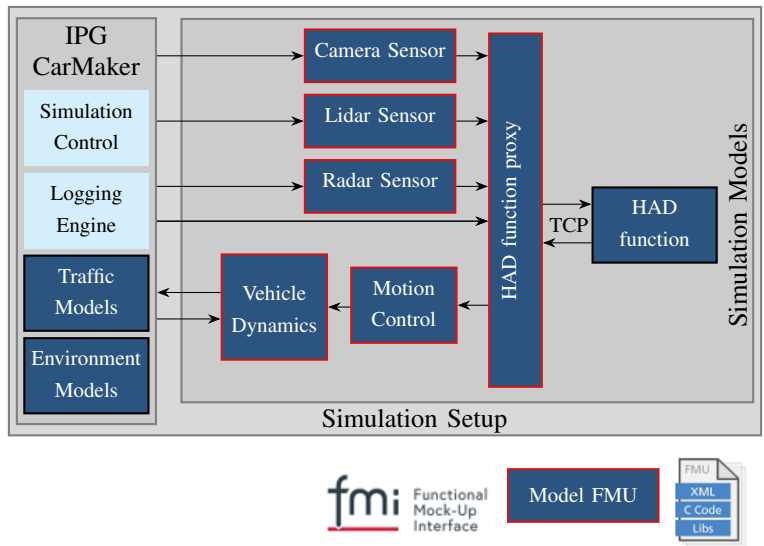}
	\end{adjustbox}
    \caption{Modular structure of the simulation environment}
    \label{fig:sim_env}
\end{figure*}
\label{subsec:simulator_setup}
The simulation tool CarMaker \footnote{CarMaker from IPG Automotive in version 8.0.2} serves as a basis for the simulator setup \cite{carmaker}. As an open integration and test platform, CarMaker provides a central control unit running the closed-loop simulation. It includes all proprietary and external models according to the given scenario and constraints. In this case, proprietary models of the complete simulation environment comprise the road, environment and traffic models. Six external models were integrated as FMUs (Functional Mock-up Unit) via an extended version of the Open Simulation Interface (OSI) \cite{osi}, realizing a setup with three sensor models (camera, lidar \cite{9548071}, and radar), an autonomous driving function (see section~\ref{sec:hadf}), a motion control model and a vehicle dynamics model. This simulator setup is shown in Fig~\ref{fig:sim_env}. The output quantities of each simulation are handed over to the optimization setup described in section~\ref{sec:opt_steup}. The next scenario to be executed is then chosen directly by the optimization setup via script commands, leading to the optimization workflow pictured in Fig~\ref{fig:optimization}.

\subsection{Driving Function}
\label{sec:hadf}
\begin{figure}[th]
    \centering
    \begin{adjustbox}{center,width=0.78\linewidth,margin*=0ex 0ex 0ex 1ex}
		\tikzsetnextfilename{had_func}
\begin{tikzpicture}[
invisibox/.style={rectangle, inner sep=5pt},
textbox/.style={rectangle,minimum height=1cm,minimum width=1.4cm,draw=black!50,fill=black!15,thick, inner sep=5pt},
background/.style={rectangle,minimum height=4cm,minimum width=6.5cm,draw=blue!50,fill=blue!15,thick, inner sep=5pt}
]
\node[background, anchor=south, align=center] (bg) {};
\node[textbox, anchor=south, align=center] at  ($(bg.north) + (down:33pt) + (left:18pt)$) (routing) {\footnotesize Routing};
\node[invisibox, anchor=east, align=center] at  ($(routing.west) + (down:0pt) + (left:80pt)$) (target) {};
\node[invisibox, anchor=east, align=center] at  ($(routing.west) + (up:0pt) + (left:58pt)$) (target2) {\footnotesize Target \\ \footnotesize Position};
\node[textbox, anchor=north, align=center] at  ($(routing.west) + (down:8pt) + (left:30pt)$) (sensorfus) {\footnotesize Sensor \\ \footnotesize Fusion};
\node[invisibox, anchor=east, align=center] at  ($(sensorfus.west) + (down:0pt) + (left:30pt)$) (sensinf) {};
\node[invisibox, anchor=east, align=center] at  ($(sensorfus.west) + (up:0pt) + (left:3pt)$) (sensinf2) {\footnotesize Sensor \\ \footnotesize Information};
\node[textbox, anchor=north, align=center] at  ($(sensorfus.south) + (down:3pt) + (right:50pt)$) (objtrack){\footnotesize Object \\ \footnotesize Tracking};
\node[textbox, anchor=west, align=center] at  ($(objtrack.east) + (up:0pt) + (right:7pt)$) (objfilter){\footnotesize Object \\ \footnotesize Filter};
\node[textbox, anchor=west, align=center] at  ($(objfilter.west) + (up:32pt) + (right:30pt)$) (trajplan){\footnotesize Trajectory \\ \footnotesize Planning};
\node[invisibox, anchor=west, align=center] at  ($(trajplan.east) + (down:0pt) + (right:30pt)$) (traject) {};
\node[invisibox, anchor=west, align=center] at  ($(trajplan.east) + (up:5pt) + (right:5pt)$) (traject2) {\footnotesize Trajectory};
\node[textbox, anchor=north, align=center] at  ($(sensorfus.south) + (down:25pt) + (right:0pt)$) (localization)
{\footnotesize Local-\\\footnotesize ization};

\node[invisibox, anchor=east, align=center] at  ($(localization.west) + (down:0pt) + (left:30pt)$) (odometry) {};
\node[invisibox, anchor=east, align=center] at  ($(localization.west) + (up:5pt) + (left:3pt)$) (odometry2) {\footnotesize Odometry};

\draw[arrows = {-Stealth[scale=1]}] (target) -- (routing);
\draw[arrows = {-Stealth[scale=1]}] (sensinf) -- (sensorfus);
\draw[arrows = {-Stealth[scale=1]}] (odometry) -- (localization);
\draw[arrows = {-Stealth[scale=1]}] (trajplan) -- (traject);
\draw[arrows = {-Stealth[scale=1]},densely dashed] (sensorfus) |- (objtrack);
\draw[arrows = {-Stealth[scale=1]},densely dashed] (sensorfus) -| (objfilter);
\draw[arrows = {-Stealth[scale=1]},densely dashed] (objtrack) -- (objfilter);
\draw[arrows = {-Stealth[scale=1]},densely dashed] (localization) -| (objtrack);
\draw[arrows = {-Stealth[scale=1]},densely dashed] (localization) -| (objfilter);
\draw[arrows = {-Stealth[scale=1]},densely dashed] (localization) -| (trajplan);
\draw[arrows = {-Stealth[scale=1]},densely dashed] (routing) -| (trajplan);
\draw[arrows = {-Stealth[scale=1]},densely dashed] (objfilter) -| (trajplan);

\end{tikzpicture}
	\end{adjustbox}
    \caption{Modular structure of the driving function}
    \label{fig:had_function}
\end{figure}
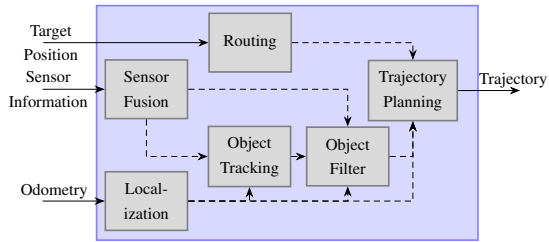

To calculate the trajectory of the ego vehicle, a lightweight and highly automated driving function is used. The function is centered around a modified, curvature-aware version of the Intelligent Driver Model (IDM) as used in \cite{zofka2016simulation}, initially introduced in \cite{treiber2000congested}. To achieve a modular system, the function is implemented using the Robot Operating System (ROS) framework \cite{quigley2009ros}. The system comprises six modules, as shown in Fig.~\ref{fig:had_function}: a sensor fusion module to join the information from the three sensors, a tracking algorithm to keep track of occluded traffic participants, a filter module to extract the relevant objects, a routing algorithm, a localization part to create an estimate of the own global position using odometry information, and finally, a trajectory module, planning a trajectory with a velocity profile.

The routing algorithm uses a high-definition map to extract the road topology. The relevant objects identified within the object filter are projected onto the path of the ego vehicle. Thereby, a distance and differential velocity can be calculated to be used within the IDM. Filtering is done considering the type of an object, as well as its position relative to the road network and the ego vehicle. Moreover, the curvature of the path is considered by converting it to a velocity limit using maximum lateral acceleration. This velocity limit is treated as a separate object for the IDM to achieve smooth cornering behavior.

A gateway architecture was used to comply with the standardized FMI/OSI interface described in section \ref{subsec:simulator_setup} while maintaining the platform's independence.
Thereby, a TCP proxy was integrated into the simulator as an FMU. The proxy forwards the messages via TCP to the communication layer of the driving function, where the messages are then converted into equivalent ROS messages. With this, the driving function can be run within a docker environment.

\subsection{Logical Scenario}
\label{sec:log_scenario}
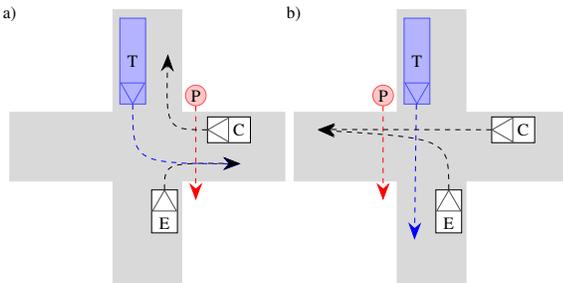
\begin{figure}[tbh]
    \centering
    \begin{adjustbox}{center,width=0.65\linewidth,margin*=-5ex 0ex 0ex 1ex}
		\tikzsetnextfilename{scenarios}
\begin{tikzpicture}[
street/.style={rectangle, minimum width=40pt, minimum height=160pt, fill=#1!30, draw=#1!30, text=black},
truck/.style={rectangle, minimum width=15pt, minimum height=50pt, fill=#1!30, draw=#1!70, text=black},
ego/.style={rectangle, minimum width=15pt, minimum height=25pt, text=black, fill=white, draw=black},
triangle/.style = {draw=#1!70, regular polygon, regular polygon sides=3, inner sep=0.1cm}
]

% scenario 1
\node[street=gray, anchor=west, align=center] (IS1) {};
\node[anchor=north, text width=95pt, align=center] at ($(IS1.north west) + (up:5pt) + (left:60pt)$) (ego1text){a)};
\node[street=gray, anchor=west, align=center, rotate=90] at  ($(IS1.center) + (down:20pt) + (left:0pt)$) (IS1b) {};

\node[circle,minimum size=12pt,inner sep=0pt] at ($(IS1.east) + (up:60pt) + (left:8pt)$) (cd1) {};
\node[circle,minimum size=12pt,inner sep=0pt] at ($(IS1b.south) + (down:10pt) + (left:20pt)$) (etd1) {};
\node[circle,minimum size=12pt,inner sep=0pt] at ($(IS1.east) + (down:37pt) + (right:8pt)$) (pd1) {};

\node[truck=blue, anchor=west, align=center] at  ($(IS1.west) + (up:50pt) + (right:4pt)$) (truck1) {T};
\node[triangle=blue, anchor=north, align=center, rotate=180] at  ($(truck1.south) + (up:0pt) + (left:0pt)$) (truck1b) {};
\node[ego, anchor=south, align=center] at  ($(IS1.south) + (up:30pt) + (right:10pt)$) (ego1) {};
\node[triangle=black, anchor=north, align=center] at  ($(ego1.north) + (up:0pt) + (left:0pt)$) (ego1b) {};
\node[anchor=north, text width=95pt, align=center] at ($(ego1b.south) + (down:0pt) + (right:0pt)$) (ego1text){E};
\node[ego, anchor=south, align=center, rotate=90] at  ($(IS1.east) + (up:10pt) + (right:40pt)$) (car1) {};
\node[triangle=black, anchor=north, align=center, rotate=90] at  ($(car1.north) + (up:0pt) + (left:0pt)$) (car1b) {};
\node[anchor=north, text width=95pt, align=center, ->] at ($(car1b.east) + (up:2pt) + (right:10pt)$) (car1text){C};
%edge[arrows = {-Stealth[scale=2]}, dashed, controls=+(right:5mm) and +(down:5mm)](cd1.south);
\node[draw,circle,minimum size=12pt,inner sep=0pt, fill=red!22, draw=red!70] at ($(IS1.east) + (up:30pt) + (right:8pt)$) (P1) {P}
edge[arrows = {-Stealth[scale=2]}, dashed, draw=red] (pd1.north);
\draw[arrows = {-Stealth[scale=2]},draw=blue,dashed] (truck1) ..controls  +(0.0,-2.1).. (etd1);
\draw[arrows = {-Stealth[scale=2]},draw,dashed] (car1) ..controls  +(-1.3,0.0).. (cd1);
\draw[arrows = {-Stealth[scale=2]},draw,dashed] (ego1) ..controls  +(0.0,1.0).. (etd1);

%scenario 3
\node[street=gray, anchor=west, align=center] at  ($(IS1.east) + (down:0pt) + (right:125pt)$) (IS3) {};
\node[anchor=north, text width=95pt, align=center] at ($(IS3.north west) + (up:5pt) + (left:60pt)$) (ego1text){b)};
\node[street=gray, anchor=west, align=center, rotate=90] at  ($(IS3.center) + (down:20pt) + (left:0pt)$) (IS3b) {};

\node[circle,minimum size=12pt,inner sep=0pt] at ($(IS3b.west) + (down:40pt) + (left:10pt)$) (td3) {};
\node[circle,minimum size=12pt,inner sep=0pt] at ($(IS3b.north) + (up:10pt) + (right:8pt)$) (ecd3) {};
\node[circle,minimum size=12pt,inner sep=0pt] at ($(IS3.west) + (down:37pt) + (left:8pt)$) (pd3) {};

\node[truck=blue, anchor=west, align=center] at  ($(IS3.west) + (up:50pt) + (right:4pt)$) (truck3) {T};
\node[triangle=blue, anchor=north, align=center, rotate=180] at  ($(truck3.south) + (up:0pt) + (left:0pt)$) (truck1b) {};
\node[ego, anchor=south, align=center] at  ($(IS3.south) + (up:30pt) + (right:10pt)$) (ego3) {};
\node[triangle=black, anchor=north, align=center] at  ($(ego3.north) + (up:0pt) + (left:0pt)$) (ego3b) {};
\node[anchor=north, text width=95pt, align=center] at ($(ego3b.south) + (down:0pt) + (right:0pt)$) (ego3text){E};
\node[ego, anchor=south, align=center, rotate=90] at  ($(IS3.east) + (up:10pt) + (right:40pt)$) (car3) {};
\node[triangle=black, anchor=north, align=center, rotate=90] at  ($(car3.north) + (up:0pt) + (left:0pt)$) (car3b) {};
\node[anchor=north, text width=95pt, align=center] at ($(car3b.east) + (up:2pt) + (right:10pt)$) (car3text){C};
\node[draw,circle,minimum size=12pt,inner sep=0pt, fill=red!22, draw=red!70] at ($(IS3.west) + (up:30pt) + (left:8pt)$) {P}
edge[arrows = {-Stealth[scale=2]}, dashed, draw=red] (pd3.north);
\draw[arrows = {-Stealth[scale=2]},draw=blue,dashed] (truck3) -- (td3);
\draw[arrows = {-Stealth[scale=2]},draw,dashed] (ego3) ..controls  +(0.0,1.5).. (ecd3);
\draw[arrows = {-Stealth[scale=2]},draw,dashed] (car3) -- (ecd3);

\end{tikzpicture}
	\end{adjustbox}
    \caption{Two experimental scenarios: a) ego turns right, b) ego turns left. C: car, E: ego vehicle, P: pedestrian, T: truck.}
    \label{fig:scenarios}
\end{figure}
As shown in Fig.~\ref{fig:scenarios}, two logical intersection scenarios are used for the experiments.
Each scenario consists of an ego vehicle (E), a pedestrian (P), a second car (C), and a truck (T).
Both scenarios vary in the actors' starting position and maneuvers.
In scenario A, the ego vehicle is turning right and, therefore, crosses the trajectories of the pedestrian and the truck but not the car's trajectory, whereas, in scenario B, the ego vehicle is turning left and crosses the trajectories of all three adversary traffic participants.
The two scenarios lead to different behaviors of the ego vehicle since it reacts to other participants blocking its route.
The following ranges were chosen as parameter ranges, for which the optimal parameter sets have to be found during the scenario exploration:
\begin{itemize}
    \item \textbf{Pedestrian delay}: The pedestrian waits for a given time $t_{P_{delay}}$ in $\SI{}{\second}$ before crossing the road, where $t_{P_{delay}} \in \{0.0, ..., 7.0\}$. $50$ samples with a step size of $\SI{0.14}{\second}$ were taken.
    \item \textbf{Ego position}: The ego vehicle starts at a given s-coordinate $s_{E_{start}}$ in $m$ along the road, where $s_{E_{start}} \in \{27.99, ...77.99\}$. $250$ samples with a step size of $\SI{0.2}{\metre}$ were taken.
    \item \textbf{Car speed}: The maximum speed $v_{C_{max}}$ that the other car is allowed to achieve in $\SI{}{\metre\per\second}$, where $v_{C_{max}} \in \{12.5,..., 30.0\}$. $50$ samples with a step size of $\SI{0.35}{\metre\per\second}$ were taken.
\end{itemize}

This setup results in $625.000$ scenarios.
If each scenario takes approximately $\SI{30}{\second}$ for simulation execution and calculation of criticality metrics, the execution of all 625.000 scenarios takes more than 217 days of non-stop simulation on a single machine.
Additionally, the simulation time grows exponentially, with scenarios getting more complex by additional parameters and parameter ranges.

\subsection{Optimization Problem}

The goal of the scenario exploration is to find all critical scenarios involving the ego vehicle and the pedestrian, the most vulnerable road user (VRU) in this scenario.
Therefore, the criticality between the ego vehicle and the pedestrian is optimized.
The criticality is measured by criticality metrics, and the optimization aims to find scenarios evaluated to what degree they contain a potentially critical situation.
Five criticality metrics were utilized as the objective function to be optimized by the Bayesian optimization:
\begin{itemize}
    \item \textbf{Euclidean Distance}: Direct distance between the center of mass of two vehicles. 
    \item \textbf{Trajectory Distance}: Distance between two traffic participants along their trajectories and road network.
    \item \textbf{Worst-time-to-collision} (WTTC): Metricetric based on time-to-collision (TTC) \cite{hayward_near_1972}, but without the TTC's limitation to car following scenarios \cite{wachenfeld2016worst}.
    \item \textbf{Gap time} (GT): The predicted distance in time between the two traffic participants crossing an intersection \cite{allen1978analysis}.
    \item \textbf{Post-encroachment-time} (PET): The actual distance in time between the two traffic participants crossing an intersection \cite{allen1978analysis}.
\end{itemize}

All metrics above require to minimize their output value to optimize the criticality within the logical scenario.

\subsection{Experiments and Results}
All experiments evaluate the criticality of the scenario regarding the ego vehicle and the pedestrian as the most VRU in this scenario.
Some scenarios might lead to critical situations between the ego vehicle and other traffic participants.
However, these scenarios are neglected.
In general, different metrics cannot be compared directly, e.g., a critical scenario in Fig~\ref{fig:exp2} b) which is indicated by a red dot is not equally critical to a red colored scenario in Fig~\ref{fig:exp2} c).

\subsubsection{Experiment 1}
\begin{figure}[tb]
    \centering
		\includegraphics[width=\linewidth]{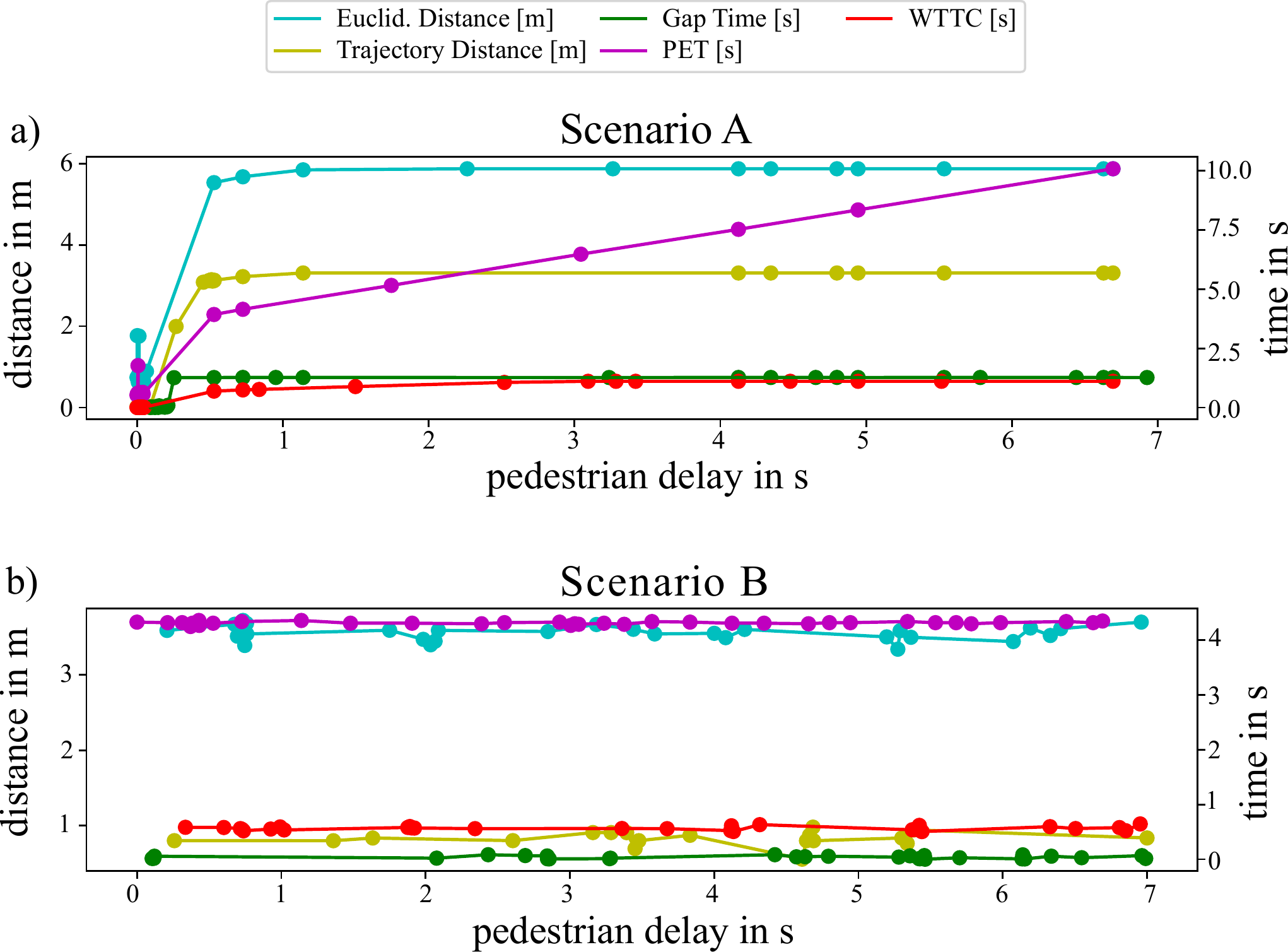}
    \caption{Bayes optimization results of both scenarios on a one-dimensional parameter space. The metrics used for optimization are Euclidean and trajectory distance, gap time, post-encroachment-time, and worst-time-to-collision.}
    \label{fig:1dresults}
\end{figure}
In the first set of experiments, only the \textbf{pedestrian delay} is varied and optimized throughout all simulations, with a set value of $s_{E_{start}} = \SI{60.0}{\metre}$ in scenario A, $s_{E_{start}} = \SI{67.0}{\metre}$ in scenario B, and $v_{C_{max}} = \SI{15.0}{\metre\per\second}$ in both scenarios.
The results are shown in Fig.~\ref{fig:1dresults} a) for scenario A and in b) for scenario B.
In scenario A, critical scenarios are found for a delay near $\SI{0.0}{\second}$, and for all metrics except PET, there are no changes in criticality for a delay over approximately $\SI{1.5}{\second}$.
Although PET values change after that, scenarios are not critical since the result is growing.
Further, \cite{allen1978analysis} set the threshold for critical scenarios to $PET < \SI{1.5}{\second}$.
The used criticality metrics in scenario B indicate no change in criticality for a varying pedestrian delay, and therefore, the pedestrian delay has no influence on the outcome of scenario B for the chosen values of the other two parameters.

\subsubsection{Experiment 2}
\begin{figure*}[t]
    \centering
		\includegraphics[width=\linewidth]{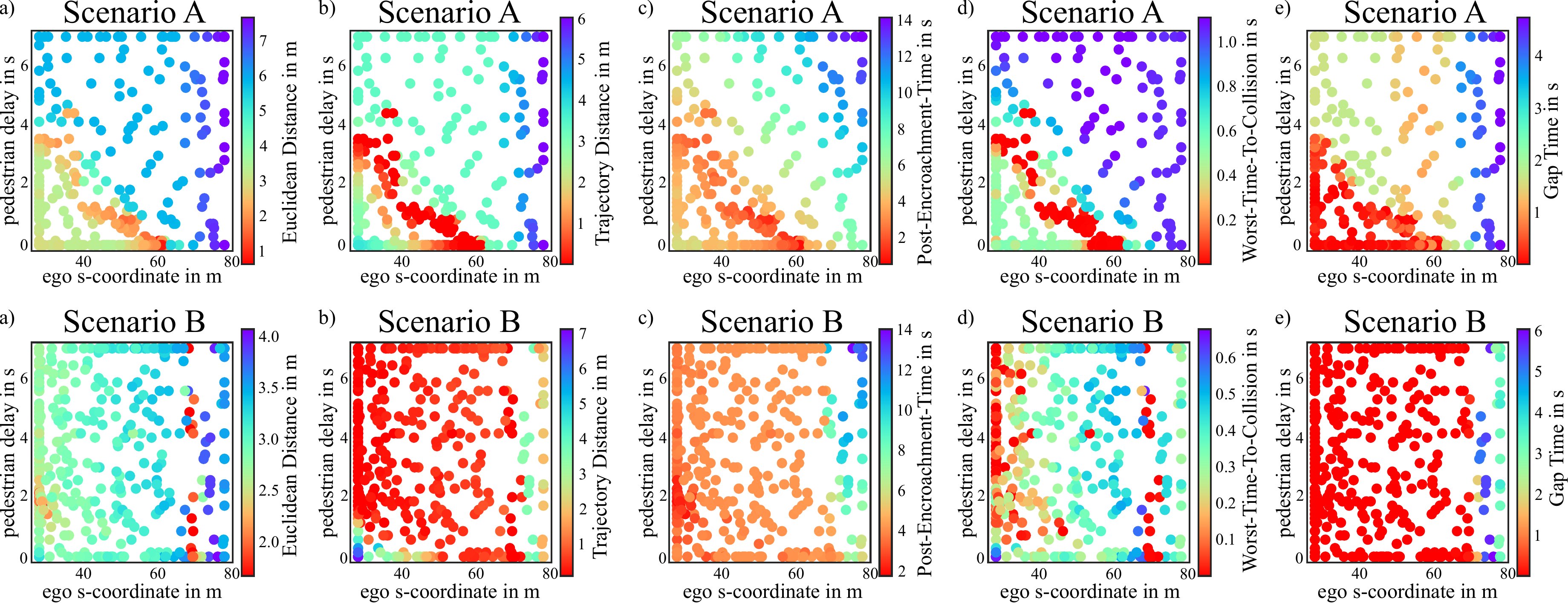}
    \caption{Bayes optimization results of both scenarios on a three-dimensional parameter space. The metrics used for optimization are Euclidean (a) and trajectory (b) distance, post-encroachment-time (c), worst-time-to-collision (d), and gap time (e) and are shown in a two-dimensional space since the variable car speed has almost no influence on the scenario outcome.}
    \label{fig:exp2}
\end{figure*}
\begin{figure}[t]
    \centering
		\includegraphics[width=\linewidth]{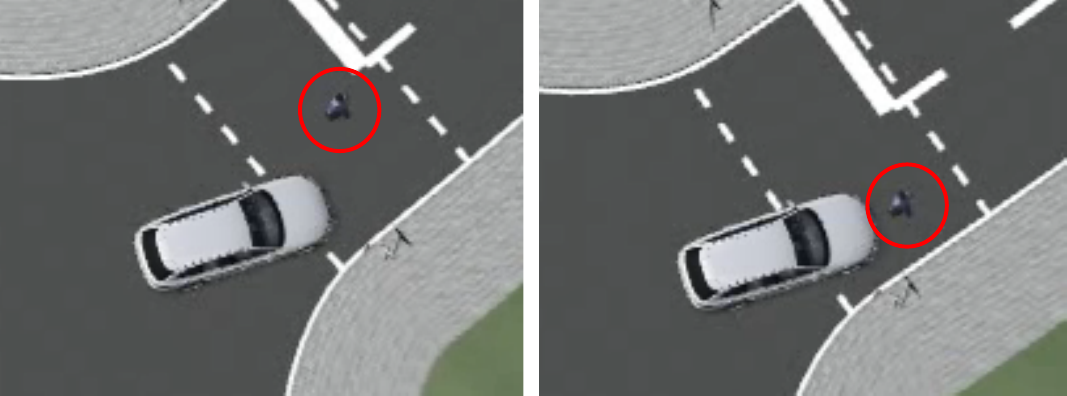}
    \caption{Screenshot from a less critical scenario (left) and critical scenario where the ego vehicle almost hits the pedestrian (right).}
    \label{fig:critexp2}
\end{figure}
In the second setup, all three parameters are varied and optimized as described in section~\ref{sec:opt_steup} and \ref{sec:log_scenario}.
The results for these experiments are shown in Fig.~\ref{fig:exp2}, where a) shows results for the Euclidean distance, b) trajectory distance, c) post-encroachment-time, d) worst-time-to-collision, and e) gap time.
In both scenarios, the \textbf{car speed} seems to have no or almost no visible influence on simulation results regarding the criticality of the pedestrian's situation.
Therefore, a three-dimensional plot can be reduced to the two dimensions of \textbf{pedestrian delay} and \textbf{ego s-coordinate}. 
However, this does not mean that there is no influence at all, and outliers or deviations in the plot that are not congruent with other values around them might be influenced by the car's speed.
An explanation for the lack of influence could be, that in scenario A, the ego vehicle's trajectory and the car's trajectory are not intersecting, and the ego has no need to react to the car.
Also, the other car is starting close to the intersection in both scenarios, it might simply not have had enough time to accelerate until it reaches the intersection and always pass the ego vehicle at the same speed.

In general, scenario A has three different outcomes regarding the criticality of the pedestrian's situation: the ego vehicle reaches the intersection and stops for the pedestrian who is crossing the street. This result can be observed in scenarios at the bottom left corner in Fig.~\ref{fig:exp2} a)-e) where the criticality decreases for most metrics.
The second result consists of the ego vehicle passing the intersection before the pedestrian or the truck, which are mostly scenarios at the top half and right half in Fig~\ref{fig:exp2} a)-e).
The last variant are scenarios where the ego vehicle reaches the intersection right after the truck and, therefore, cannot see the pedestrian. 
These scenarios are the critical scenarios around the diagonal near the bottom left corner in Fig.~\ref{fig:exp2} a)-e).

In scenario B, the ego vehicle not only reacts to the pedestrian and the truck but also the other car.
The outcomes of scenario B are the same as in scenario A. 
However, they are distributed differently over the parameter space:
the less critical area in the bottom left corner, which can be clearly observed in Fig.~\ref{fig:exp2} b) and d) results from a parameter set, where the ego vehicle intersects the pedestrian's path after they crossed the road.
The left part of Fig.~\ref{fig:critexp2} shows a screenshot of this situation taken during the simulation.
On the right side, where the s-coordinate has its highest value, the ego vehicle passes the intersection before the pedestrian, followed by a critical line around $\SI{70}{\metre}$ with near-collisions.
In the middle and left part of the s-coordinate, the ego vehicle waits for the pedestrian to pass, and the critical cluster at delay $\SI{2}{\second}$ results from interference with the truck.

The comparison of the results of both scenarios leads to the following conclusions:
\begin{itemize}
    \item Scenario A has a higher variance in criticality than scenario~B,
    \item car speed has no recognizable influence in scenario A and almost no influence in B, 
    \item some metrics are more sensitive, e.g., trajectory distance and gap time, and
    \item some metrics lead to similar patterns in criticality, e.g., bottom left corner in scenario~A or scenarios with a small ego vehicle s-coordinate in scenario~B.
\end{itemize}

\subsubsection{Experiment 3}
\begin{figure}[t]
    \centering
		\includegraphics[width=\linewidth]{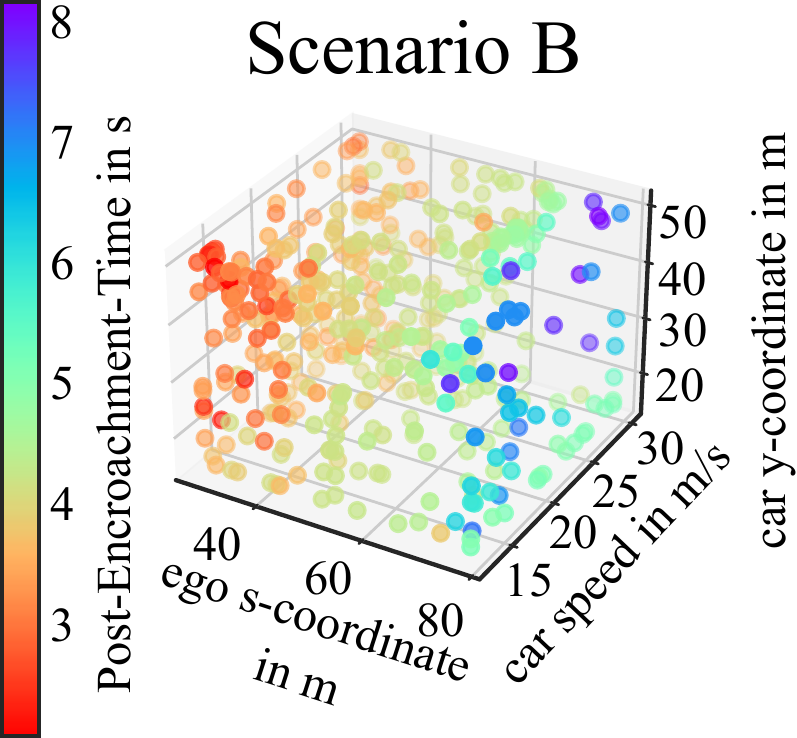}
    \caption{Exemplary Bayes optimization results of scenario B on a three-dimensional parameter space for the metric post-encroachment-time.}
    \label{fig:exp3}
\end{figure}
In the third setup, the pedestrian delay $t_{P_{delay}}$ is replaced by a new variable $y_{C_{start}}$ to see if the car speed has more influence on the scenario outcome:
\begin{itemize}
    \item \textbf{Car position}: The y-coordinate $y_{C_{start}}$ in $\SI{}{\metre}$ fulfils $y_{C_{start}} \in \{15.0,..., 50.0\}$. $50$ samples with a step size of $\SI{0.7}{\metre}$ were taken.
\end{itemize}
All three variables are varied and optimized, and the five previously mentioned metrics are used.
The results for scenario A do not show any influence of the car's speed or y position on the outcome.
This is not surprising since there is no trajectory intersection between the ego vehicle and the car.
The results show, the outcome only depends on the s-coordinate.
As Fig.~\ref{fig:exp3} shows, in scenario B the ego s-coordinate still has the most influence on the outcome. 
However, in some areas in the parameter space car speed and y-coordinate also affect the scenario criticality.

\subsection{Results}

In our experiments, we could show that even though different metrics were used, they led to similar critical scenario clusters although, these metrics are not comparable in the severity of the measured criticality and their sensitivity.
Moreover, our approach led to a reduction of the amount of necessary simulation: instead of executing more than half a million scenarios, only about 430 were executed in experiments 2 and 3, respectively.
Additionally, we were able to show that the variable \textit{car speed} has no influence in scenario A and can be neglected to reduce the number of scenarios or replaced by another variable with more influence.
\section{\uppercase{Conclusion and Future Work}}
\label{sec:conclusion}

In this work, we used an optimization algorithm to find critical scenarios for the develoing and testing of automated and highly automated driving systems.
Bayes optimization with Gaussian process was utilized in combination with five criticality metrics from the automotive domain to calculate the process output.
This approach was used in two different experiments and evaluated accordingly.

Derived from the evaluation of the experiments and the results of this work, additional questions arise.
Results of the same scenario, models, criticality metrics, and the same driving function could be used by different simulation tools and compared regarding their deviation.
Furthermore, it is harder to measure criticality for other scenarios, i.e., scenarios with more traffic participants.
Our experiments only focused on criticality metrics between the ego vehicle and the most VRU, the pedestrian.
However, such a choice might not always be obvious or changing during one scenario, e.g., in urban rush hour traffic with a high density of traffic participants.
Metrics to evaluate the relation between the ego and more than one adversary traffic participant or the whole scenario situation are needed to make more objective conclusions.
In future work, the problem of finding more objective metrics for scenarios will be approached to be able to find critical situations between the ego vehicle and the sum of all other traffic participants.

\section*{\uppercase{Acknowledgements}}
This research is funded by the ‘‘Simulationsbasiertes Entwickeln und Testen von automatisiertem Fahren (SET Level)-Simulation-Based Development and Testing of Automated Driving,’’ a successor project to the project ‘‘Projekt zur Etablierung von generell akzeptierten Gütekriterien, Werkzeugen und Methoden sowie Szenarien und Situationen zur Freigabe hochautomatisierter Fahrfunktionen (PEGASUS)’’ and a Project in the PEGASUS Family, promoted by the German Federal Ministry for Economic Affairs and Energy (BMWi) under the grant number 19 A 19004.

\bibliographystyle{apalike}
{\small
\bibliography{paper}}

\begin{thebibliography}{}

\bibitem[Abeysirigoonawardena et~al., 2019]{abeysirigoonawardena2019generating}
Abeysirigoonawardena, Y., Shkurti, F., and Dudek, G. (2019).
\newblock Generating adversarial driving scenarios in high-fidelity simulators.
\newblock In {\em 2019 International Conference on Robotics and Automation
  (ICRA)}, pages 8271--8277. IEEE.

\bibitem[Allen et~al., 1978]{allen1978analysis}
Allen, B.~L., Shin, B.~T., and Cooper, P.~J. (1978).
\newblock Analysis of traffic conflicts and collisions.
\newblock Technical report.

\bibitem[{ASAM OSI}, 2021]{osi}
{ASAM OSI} (2021).
\newblock {ASAM OSI® (Open Simulation Interface)}.
\newblock Accessed: Dec. 15, 2021.

\bibitem[Bach et~al., 2016]{bach2016model}
Bach, J., Otten, S., and Sax, E. (2016).
\newblock Model based scenario specification for development and test of
  automated driving functions.
\newblock In {\em 2016 IEEE Intelligent Vehicles Symposium (IV)}, pages
  1149--1155. IEEE.

\bibitem[Baumann et~al., 2021]{baumann2021automatic}
Baumann, D., Pfeffer, R., and Sax, E. (2021).
\newblock Automatic generation of critical test cases for the development of
  highly automated driving functions.
\newblock In {\em 2021 IEEE 93rd Vehicular Technology Conference
  (VTC2021-Spring)}, pages 1--5. IEEE.

\bibitem[Bussler et~al., 2020]{bussler2020application}
Bussler, A., Hartjen, L., Philipp, R., and Schuldt, F. (2020).
\newblock Application of evolutionary algorithms and criticality metrics for
  the verification and validation of automated driving systems at urban
  intersections.
\newblock In {\em 2020 IEEE Intelligent Vehicles Symposium (IV)}, pages
  128--135. IEEE.

\bibitem[Dosovitskiy et~al., 2017]{Dosovitskiy17}
Dosovitskiy, A., Ros, G., Codevilla, F., Lopez, A., and Koltun, V. (2017).
\newblock {CARLA}: {An} open urban driving simulator.
\newblock In {\em Proceedings of the 1st Annual Conference on Robot Learning},
  pages 1--16.

\bibitem[{dSPACE}, 2021]{dSPACE}
{dSPACE} (2021).
\newblock {SIMPHERA - Web-based solution for simulation and validation in
  autonomous driving development}.
\newblock Accessed: Jan. 19, 2022.

\bibitem[Gartner, 2021]{Gartner}
Gartner (2021).
\newblock Top strategic technology trends for 2022.
\newblock
  https://www.gartner.com/en/information-technology/insights/top-technology-trends.
\newblock Accessed: 2021-10-25.

\bibitem[Greenhill et~al., 2020]{bayes_opt_review}
Greenhill, S., Rana, S., Gupta, S., Vellanki, P., and Venkatesh, S. (2020).
\newblock Bayesian optimization for adaptive experimental design: A review.
\newblock {\em IEEE Access}, 8:13937--13948.

\bibitem[Hayward, 1972]{hayward_near_1972}
Hayward, J.~C. (1972).
\newblock Near miss determination through use of a scale of danger.
\newblock In {\em Unknown}.
\newblock Publisher: Pennsylvania State University University Park.

\bibitem[{IPG Automotive GmbH}, 2021]{carmaker}
{IPG Automotive GmbH} (2021).
\newblock {CarMaker}.
\newblock Accessed: Dec. 20, 2021.

\bibitem[King et~al., 2021]{king2021capturing}
King, C., Braun, T., Braess, C., Langner, J., and Sax, E. (2021).
\newblock Capturing the variety of urban logical scenarios from bird-view
  trajectories.
\newblock In {\em VEHITS}, pages 471--480.

\bibitem[Linnhoff et~al., 2021]{9548071}
Linnhoff, C., Rosenberger, P., and Winner, H. (2021).
\newblock Refining object-based lidar sensor modeling — challenging ray
  tracing as the magic bullet.
\newblock {\em IEEE Sensors Journal}, 21(21):24238--24245.

\bibitem[Lopez et~al., 2018]{SUMO2018}
Lopez, P.~A., Behrisch, M., Bieker-Walz, L., Erdmann, J., Fl{\"o}tter{\"o}d,
  Y.-P., Hilbrich, R., L{\"u}cken, L., Rummel, J., Wagner, P., and Wie{\ss}ner,
  E. (2018).
\newblock Microscopic traffic simulation using sumo.
\newblock In {\em The 21st IEEE International Conference on Intelligent
  Transportation Systems}. IEEE.

\bibitem[Menzel et~al., 2018]{menzel2018scenarios}
Menzel, T., Bagschik, G., and Maurer, M. (2018).
\newblock Scenarios for development, test and validation of automated vehicles.
\newblock In {\em 2018 IEEE Intelligent Vehicles Symp. (IV)}, pages 1821--1827.
  IEEE.

\bibitem[Otten et~al., 2018]{otten2018automated}
Otten, S., Bach, J., Wohlfahrt, C., King, C., Lier, J., Schmid, H., Schmerler,
  S., and Sax, E. (2018).
\newblock Automated assessment and evaluation of digital test drives.
\newblock In {\em Advanced Microsystems for Automotive Applications 2017},
  pages 189--199. Springer.

\bibitem[Quigley et~al., 2009]{quigley2009ros}
Quigley, M., Conley, K., Gerkey, B., Faust, J., Foote, T., Leibs, J., Wheeler,
  R., Ng, A.~Y., et~al. (2009).
\newblock Ros: an open-source robot operating system.
\newblock In {\em ICRA workshop on open source software}, volume~3, page~5.
  Kobe, Japan.

\bibitem[Sch{\"o}nemann et~al., 2018]{schonemann2018scenario}
Sch{\"o}nemann, V., Winner, H., Glock, T., Otten, S., Sax, E., Boeddeker, B.,
  Verhaeg, G., Tronci, F., and Padilla, G.~G. (2018).
\newblock Scenario-based functional safety for automated driving on the example
  of valet parking.
\newblock In {\em Future of Information and Communication Conference}, pages
  53--64. Springer.

\bibitem[Sch{\"u}tt et~al., 2021]{schutt2021taxonomy}
Sch{\"u}tt, B., Steimle, M., Kramer, B., Behnecke, D., and Sax, E. (2021).
\newblock A taxonomy for quality in simulation-based development and testing of
  automated driving systems.
\newblock {\em arXiv preprint arXiv:2102.06588}.

\bibitem[Steimle et~al., 2021]{steimle2021consistent}
Steimle, M., Menzel, T., and Maurer, M. (2021).
\newblock Towards a consistent terminology for scenario-based development and
  test approaches for automated vehicles: A proposal for a structuring
  framework, a basic vocabulary, and its application.
\newblock {\em arXiv preprint arXiv:2104.09097}.

\bibitem[Treiber et~al., 2000]{treiber2000congested}
Treiber, M., Hennecke, A., and Helbing, D. (2000).
\newblock Congested traffic states in empirical observations and microscopic
  simulations.
\newblock {\em Phys. Rev. E}, 62:1805--1824.

\bibitem[Ueno et~al., 2016]{ueno2016combo}
Ueno, T., Rhone, T.~D., Hou, Z., Mizoguchi, T., and Tsuda, K. (2016).
\newblock Combo: an efficient bayesian optimization library for materials
  science.
\newblock {\em Materials discovery}, 4:18--21.

\bibitem[Wachenfeld et~al., 2016]{wachenfeld2016worst}
Wachenfeld, W., Junietz, P., Wenzel, R., and Winner, H. (2016).
\newblock The worst-time-to-collision metric for situation identification.
\newblock In {\em 2016 IEEE Intelligent Vehicles Symposium (IV)}, pages
  729--734. IEEE.

\bibitem[Zofka et~al., 2015]{zofka2015data}
Zofka, M.~R., Kuhnt, F., Kohlhaas, R., Rist, C., Schamm, T., and Z{\"o}llner,
  J.~M. (2015).
\newblock Data-driven simulation and parametrization of traffic scenarios for
  the development of advanced driver assistance systems.
\newblock In {\em 2015 18th International Conference on Information Fusion
  (Fusion)}, pages 1422--1428. IEEE.

\bibitem[Zofka et~al., 2016]{zofka2016simulation}
Zofka, M.~R., Kuhnt, F., Kohlhaas, R., and Zöllner, J.~M. (2016).
\newblock Simulation framework for the development of autonomous small scale
  vehicles.
\newblock In {\em 2016 IEEE International Conference on Simulation, Modeling,
  and Programming for Autonomous Robots, SIMPAR 2016, San Francisco, CA, USA,
  December 13-16, 2016}, pages 318--324.

\end{thebibliography}
\end{document}